\documentclass[a4paper,11pt]{article}
\pdfoutput=1 

\usepackage{jinstpub} 
\usepackage{ulem} 

\title{Ultra-clean radon-free four cylinder magnetically-coupled piston pump}

\author[a,1]{D. Schulte,\note{Corresponding author}}
\author[a]{M. Murra,}
\author[a]{P. Schulte,}
\author[a]{C. Huhmann,}
\author[a]{C. Weinheimer}

\affiliation[a]{Westfälische Wilhelms-Universität Münster, Germany}

\emailAdd{DennySchulte@uni-muenster.de}

\abstract{A high performance gas displacement pump based on four individual cylinders was developed. The magnetically-coupled pistons and the hermetically-sealed housings as well as the special cleanliness in terms of out-gassing and radio-purity make the four cylinder pump interesting for the usage in low background experiments dealing with noble gases as target. An optimized inner flow path as well as reduced dead volume with respect to the prototype single cylinder pump of the XENON1T experiment combined with advanced polymer piston gaskets and a piston coupling force of ($3177 \pm 162$)\,N ensures a high and reliable performance. Furthermore, a phase-shifted synchronization of the pistons' movement creates an additional performance boost as well as minimizes fluctuations of flow and compression. The pump is able to generate xenon flows of 474\,slpm with a pressure difference of 1.80\,bar at an inlet pressure of 2.1\,bar.}


\arxivnumber{} 

\proceeding{N$^{\text{th}}$ Workshop on X\\
  when\\
  where}

\begin{document}
\maketitle
\flushbottom

\section{Introduction}
Noble gases are deployed as target material for many rare event searches \cite{a,b,bb,bbb,c,cc,d,dd,ddd,dddd,ddddd}. Such detectors make use of the ionization and scintillation created by electron or nuclear recoils after collisions with incoming particles. In order to achieve higher sensitivities, target masses are increasing and backgrounds become more and more important. Gas and liquid purification systems are required to remove electronegative impurities produced by out-gassing from detector components, as well as intrinsic contamination by radioactive impurities. For this purpose, a prototype custom-made ultra-clean pump was developed \cite{e} for the dark matter experiment XENON1T and the double beta decay experiment nEXO on basis of the EXO-200 pump \cite{f}. This single cylinder pump was successfully installed in XENON1T as main pump recirculating the xenon through hot metal getters \cite{ff}. Commercially available pumps cannot fulfill the strict requirements in terms of radio-purity, gas containment and reliability \cite{e,f}.

A special challenge for rare event searches is the background produced by the isotope $^{222}$Rn continuously emanated from detector materials by diffusion from the $^{226}$Ra decay into the detection volume. Its half life of 3.8\,d results in a homogeneous distribution within the target noble gas where beta decays of the daughter nuclei, e.g. $^{214}$Pb, significantly increase the background rate in the region of interest. Therefore, extensive screening of the components and thus, careful material selection keeps the radon emanation rate as low as possible \cite{ff,fff,ffff}. Further reduction can be achieved by an online radon removal system based on adsorption via gas chromatography \cite{g,h} or based on cryogenic distillation \cite{i,j}.

Due to the continuous emanation, such radon removal systems need to continuously purify the detector's target mass on a faster time scale than the radon decay to reach sufficiently large reduction factors. In the case of XENONnT, a cryogenic distillation column was built. Its large throughput of 200\,slpm xenon allows to exchange the xenon inventory every 5.5\,d reaching a factor 2 radon reduction for sources emanating directly into the detector volume \cite{j}.

The system's design with liquid xenon inlet and outlet is based on novel heat exchangers featuring high xenon liquefaction capabilities \cite{jj}. In combination with the four cylinder magnetically-coupled piston pump presented in this work an energy-efficient liquid output of radon-depleted xenon is provided.
The individual cylinders of this pump are based on the prototype pump \cite{e}, but were optimized to meet current and future needs of purity and performance.

This paper is structured as follows: Section\,\ref{I} introduces the design aspects of the new four cylinder piston pump with the focus on improvements compared to the prototype of XENON1T. The realization of the synchronization of the four parallel connected cylinders, its configuration and its control interface are shown in section\,\ref{II}. The system is characterized in section\,\ref{III} including stand-alone performance measurements of each cylinder pump and the performance with different configurations of a multi-piston pump. At the end, the first long-term operation of the pump through the cryogenic distillation column is presented followed by a conclusion.

\section{Design aspects}\label{I}
The four cylinder magnetically-coupled piston pump is based on the prototype developed for the application at the experiments XENON1T and nEXO \cite{e}. This prototype was first installed and operated at the gas purification system of XENON1T \cite{ff}. Its special magnetic arrangement, where a piston housing three ring magnets with alternating longitudinal orientation is directly coupled to an external array off three magnetic rings with opposite orientation compared to the piston, allowed for a high performance xenon flow of 100\,slpm with an output differential pressure of 1.4\,bar. Furthermore, its low radon emanation of ($330 \pm 60$)\,$\upmu$Bq makes it interesting for the usage in the radon removal system of XENONnT, where a radon-free xenon gas compressor is required. However, the prototype pump performance does not meet the desired differential pressures of $\Delta p \approx 1.5$\,bar between inlet and outlet combined with a gaseous xenon flow of $Q \approx 200$\,slpm for the novel radon removal system  \cite{j}.

Therefore, the design of the cylinders was optimized. One key aspect for a good performance is the coupling force $F$ of the piston. Assuming a constant magnetic coupling force per unit length between the permanent neodymium magnets of the outer ring and the piston of the pump the magnetic coupling force $F$ scales with the circumference $U$ and thus, with the radius $r$ of the cylinder:
\begin{align}
  \label{eq:fource}
    F\sim U \sim r.
\end{align}
The differential pressure $\Delta p$ depends on the force $F$ along the piston surface area $A$:
\begin{align}
  \label{eq:pressure_difference}
    \Delta p=\frac{F}{A} \sim \frac{r}{\pi \cdot r^2}\sim \frac{1}{r}.
\end{align}
Therefore, to slightly increase the differential pressure $\Delta p$ the tube radius was reduced from 63\,mm to 50\,mm.

\begin{figure}[htbp]
\centering 
\includegraphics[width=.7\textwidth]{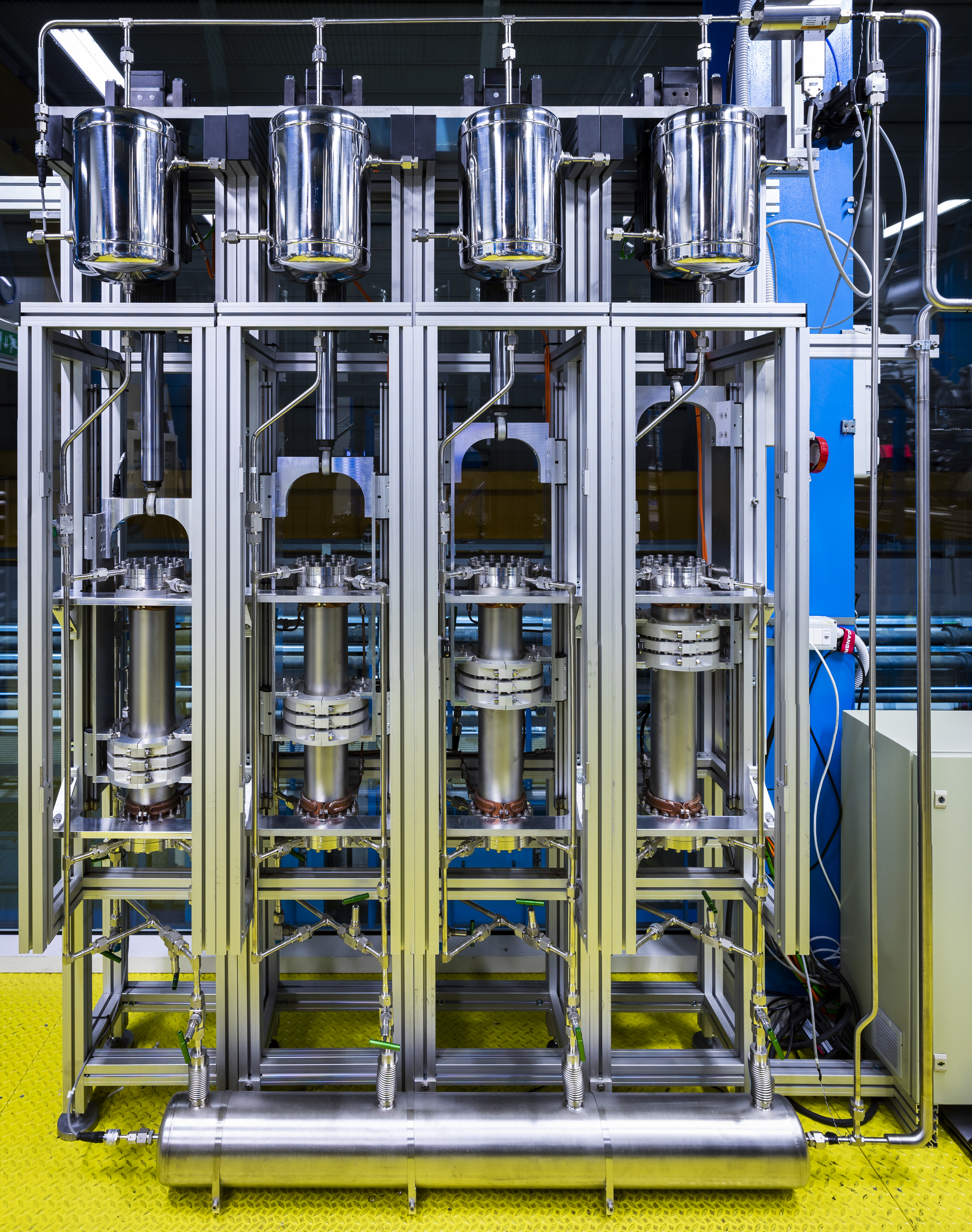}

\caption{\label{fig:1}Four magnetically-coupled piston pumps are connected in parallel to a four cylinder pump, here installed underground at XENONnT. Each pump is operated via a linear drive moving an external magnetic ring coupled to an inner magnetic piston. At the inlets, a collective buffer volume supplies each pump with gas (bottom). Each individual outlet consists of a buffer volume (top) before merging the lines to a collective outlet.}
\end{figure}

On the other hand, the flow $Q$ depends on the
effective volume per stroke and thus, on the piston surface area $A$:
\begin{align}
  \label{eq:flow}
    Q\sim A \sim r^2.
\end{align}
In order to compensate the expected lower flow $Q$ due to the reduced radius $r$, four of such cylinder pumps connected in parallel were built as visualized in figure\,\ref{fig:1}.

\newpage
The pump cylinders were constructed of monolithic type 316 L (1.4435) stainless-steel tubes. The increased nickel concentration of this steel alloy compared to other substances leads to a low permeability desired for the operation with strong magnets. Beside the above mentioned decreased tube radius, its length was increased from 530\,mm to 550\,mm creating an effective stroke volume of 3\,l.

Both ends of the tube are closed with two custom type 316 LN (1.4429) ConFlat flanges each guiding the fluid in and out. Fluid flow simulations with COMSOL Multiphysics were performed to optimize the flow path with sub-millimeter thin spring-steel flapper valves through the end caps on both sides of the cylinder. This way, the flow resistance within the pump could be reduced creating a performance boost in addition to the smaller diameter. Both, the volume above and below the piston are utilized for double stroke operation.

\begin{sloppypar}The piston inside the cylinder contains three ring magnets made of high-energy neodymium 44H, each with a length of 20\,mm, an inner diameter of 50\,mm and an outer diameter of 93\,mm. The axially magnetic-oriented rings are stacked with repulsive orientation with a 10\,mm aluminum spacer in between similar as in \cite{e}.
In order to ensure a total hermetical containment of the magnetic arrangement from the gas, the stainless-steel piston container was laser-welded. The piston container itself is sealed against the pump body with custom-made gaskets of an advanced friction bearing polymer. Several polymers were tested with sufficient properties in terms of out-gassing, radon emanation, thermal expansion, heat distortion, dynamic sliding friction and wear rate. The most promising piston gasket material was IGLIDUR A180. The usage of only clean and radio-pure components leads to a measured average radon emanation of ($75 \pm 13$)\,$\upmu$Bq for a single cylinder pump \cite{z}. Furthermore, the piston end caps contain aluminum frames to reduce the dead volume within the pump body during operation. This increased the piston height to 175\,mm.

The piston is magnetically-coupled to a stack of three external magnetic rings as in \cite{e}. Each outer magnetic ring contains 30 cylindrical oriented 20\,mm x 20\,mm x 10\,mm bar magnets made of high-energy neodymium N52 with diametrical magnetic orientation. The gap between inner magnets and outer magnets could be reduced from 8.5\,mm to 6.5\,mm compared to the prototype leading to an additional boost.\end{sloppypar}

\begin{figure}[htbp]
\centering 
\includegraphics[width=.9\textwidth]{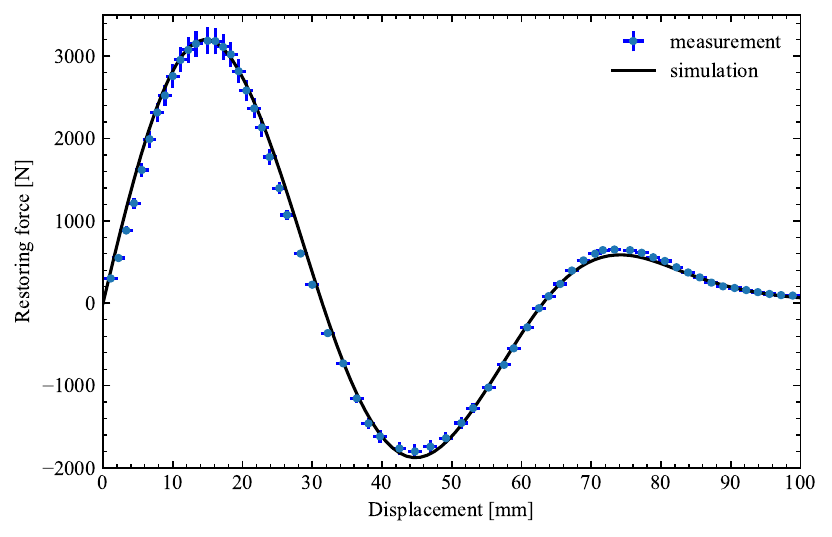}

\caption{\label{fig:2}Restoring force of one piston pump (MP502). Measurement with a load cell where the piston was moved through the fixed external ring (blue) and simulation (black). The specific shape of the curve is created by the special magnetic cross coupling arrangement.}
\end{figure}

Simulations of the restoring force show a maximum coupling for the piston of 3201\,N in agreement with the measured value of (3177$\pm$162)\,N at the cylinder pump MP502 (fig.\,\ref{fig:2}).

The maximum coupling strength of the other three pistons to their external ring were measured to be comparable as expected from the identical design of each pump.

From the coupling force $F$, the expected pressure difference $\Delta p$ across the piston can be calculated to be 
\begin{align}
\Delta p = \frac{F}{A} = 4\,\text{bar}.
\end{align}

In order to lower oscillations created by the double stroke engine combined with operation frequencies of around 1\,double-stroke per second of each cylinder pump, stainless-steel buffer volumes were installed at the inlet and at the outlet of the system. A 24\,l inlet buffer volume supplies all four cylinders with gas. On the other hand, each single piston pump contains a 12\,l buffer volume at the outlet smoothing the out-coming gas flow before the outlets are merged to the collective outlet line. To further lower fluctuations of inlet and outlet and thus, to ensure a stable operation of the connected system, the individual piston movements were synchronized to each other.

\section{Synchronization}\label{II}
\begin{figure}[htbp]
\centering 
\includegraphics[width=.8\textwidth]{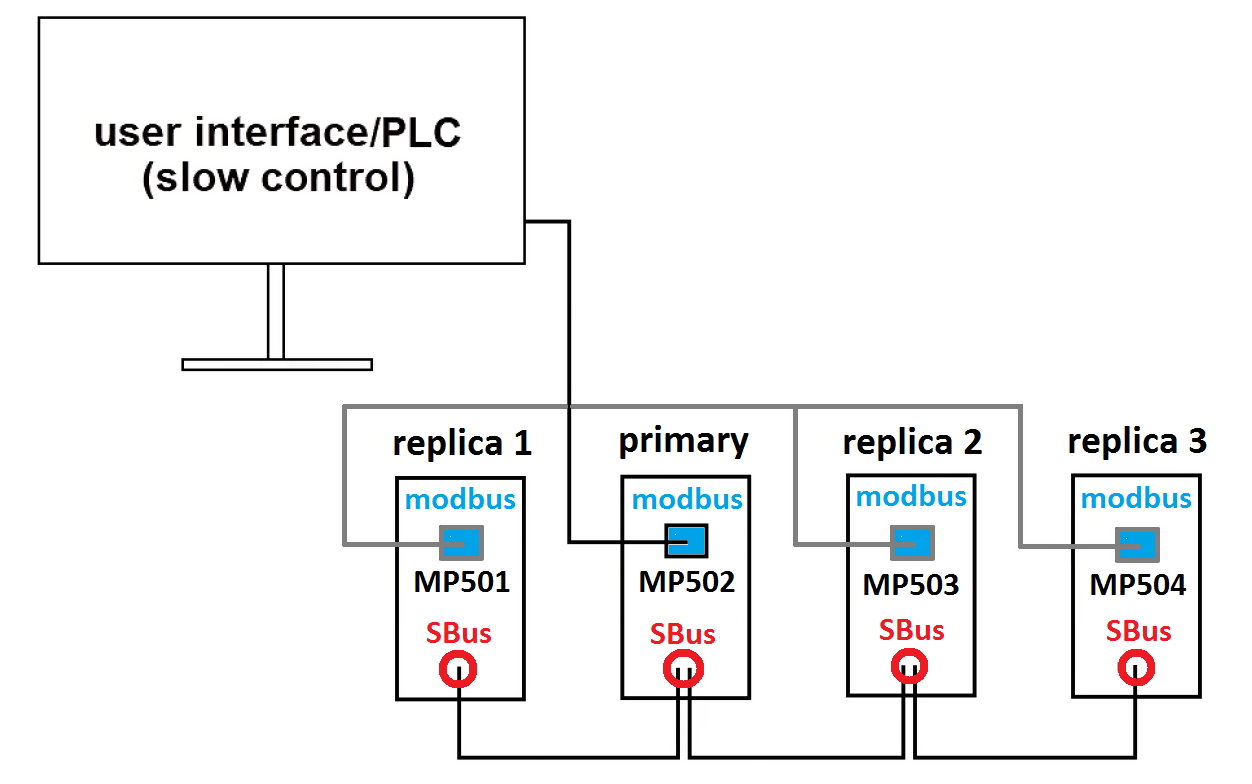}

\caption{\label{fig:3}Sketch of the communication channels between user interface and the four frequency converters realizing the synchronization. The user interface is connected to each converter via the Modbus protocol, such that each converter can be set to the primary or replica. The converters communicate to each other via SBus communication. This way status and controlling parameters are transferred.}
\end{figure}

The external ring movement of each cylinder and thus, the coupled piston, is realized with an electric cylinder drive (SEW, CMS). Similar to the prototype pump of XENON1T, a drive profile in the SEW IPOS positioning and sequence control software was programmed and uploaded to the frequency converter (SEW, MDX61B) to move the piston over the stroke length inside the pump body. In case of the four cylinder pump, the communication concept is based on a primary and replica structure as sketched in figure\,\ref{fig:3}. Here, the primary provides the position for the replica. This way, the user controls only the primary piston pump via the related converter, and the replica follow.

In order to have the possibility of changing the primary, all four converters with individual IP address are connected via the Modbus protocol to the user interface. This way, each individual pump (MP501, MP502, MP503 and MP504) can be configured as primary or replica. For the replica, a desired position shift from the primary position with regard to the stroke length of the cylinder can be selected. Here, +25\%, -25\%, -50\% (4 pumps operation), +33\% and -33\% (3 pumps operation) and -50\% (2 pumps operation) as the optimal replica shift configuration can be applied. The related phase shift ensures continuous suction and compression. The communication between primary and replica is realized via SBus channels to ensure fast and reliable communication and thus, a position synchronization.

\begin{figure}[htbp]
\centering 
\includegraphics[width=1\textwidth]{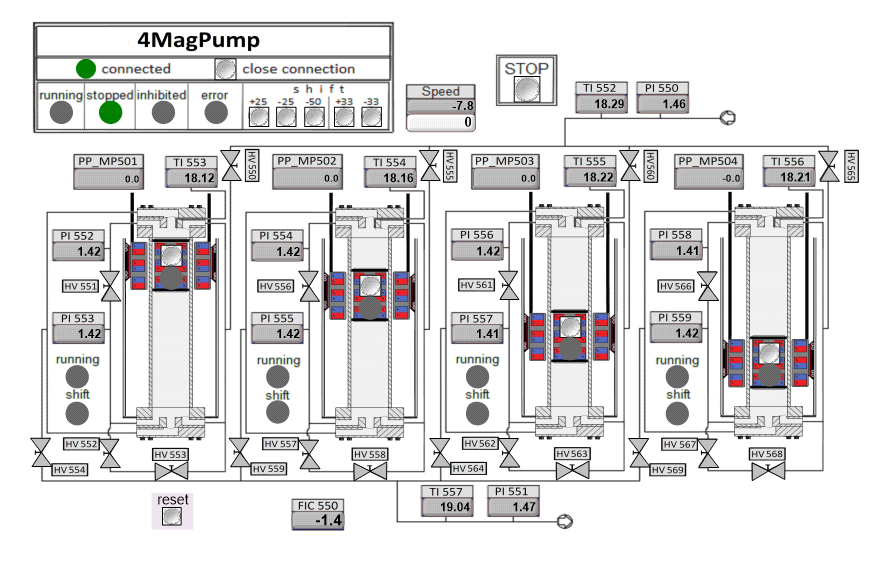}
\caption{\label{fig:4} User interface to configure and monitor the four cylinder magnetically-coupled piston pump via a Modbus connection from a PLC. Involved gas line connections and valves are visualized. Each pump can be selected to configure the shift for a replica or to set a speed for the master. Indicators of pressure, temperature and gas flow meter sensors allow for monitoring the full system. Several circular indicators show the status of the primary and the replica.}
\end{figure}

Figure\,\ref{fig:4} shows the user interface developed for the communication with a programmable logic controller (PLC). The four individual pumps connected in parallel via gas lines are visualized.

All pumps can be isolated from each other via manual valves allowing an operation with two, three or four pumps. This way the throughput can be adapted and maintenance work is possible without stopping the distillation operation.

\newpage
\section{Performance}\label{III}
In order to characterize the four cylinder pump performance, the stand-alone performance of each individual cylinder pump was measured before with xenon gas. For that, the pumps not in use are isolated from the gas system via manual valves and the pump to be characterized is operated in a closed loop of the four-cylinder pump system connecting the outlet with the inlet via a bypass valve. This way, flow and differential pressure measurements with xenon could be made at several inlet pressures by increasing the amount of gas inside the system.

\begin{figure}[htbp]
\centering 
\includegraphics[width=1\textwidth]{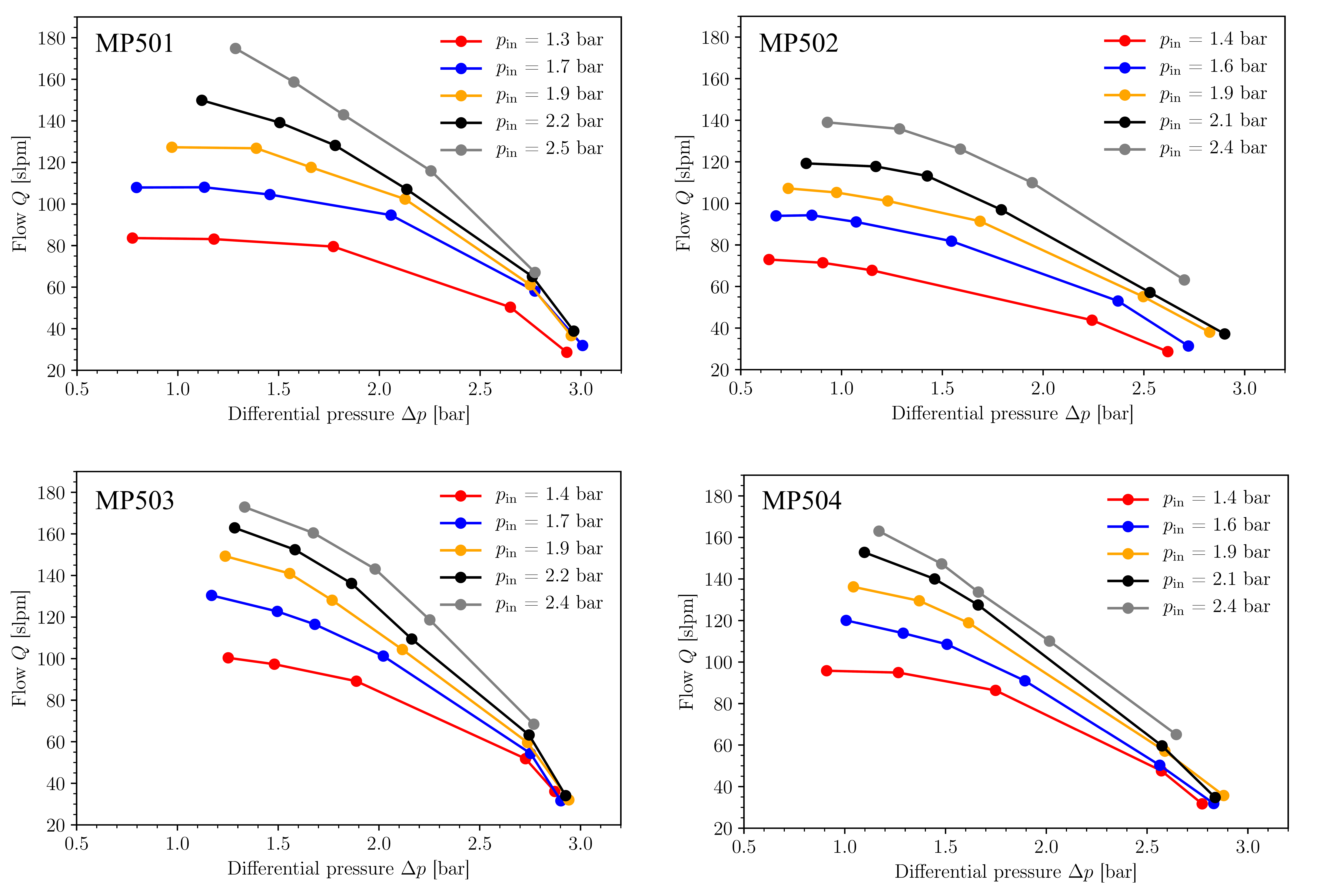}
\caption{\label{fig:5}Single performance of each cylinder pump as xenon flow versus differential pressure $\Delta p$ at several inlet pressures $p_{\mathrm{in}}$. On the top, the performances of MP501 (left) and MP502 (right) are shown. The bottom plots illustrate the results for MP503 (left) and MP504 (right). Differences in performance of the identically constructed pumps can be explained by different piston gasket sizes.}
\end{figure}

Figure\,\ref{fig:5} summarizes these measurements of flow $Q$ versus differential pressure $\Delta p$, which is the difference of the pumps outlet pressure and the pumps inlet pressure ($p_{\mathrm{out}}-p_{\mathrm{in}}$), at several inlet pressures $p_{\mathrm{in}}$ for each individual pump (MP501, MP502, MP503 and MP504).

At a differential pressure of 1.5\,bar required for the operation of the radon removal system each pump reaches flows higher 100\,slpm. Therefore, the radon removal design flow of 200\,slpm cannot be performed by a single cylinder pump, but large flows greater 200\,slpm can be expected from the four cylinder pump.

The small differences in single cylinder performance can be explained by different sizes of the piston gaskets. All four pumps demand a different size for the same performance due to small different inner diameters after honing of the cylinders. It was found that the piston gaskets need to be machined with optimal size in sub-millimeter precision. For instance, the performance of MP502 shows that gaskets small in radius lead to larger internal leakage between both sides of the piston. This results to lower flows and differential pressures. On the other hand, gaskets with larger radius than optimal increase the wear and thus, result in a decoupling of the inner piston from the external ring.

\begin{figure}[htbp]
\centering 
\includegraphics[width=0.8\textwidth]{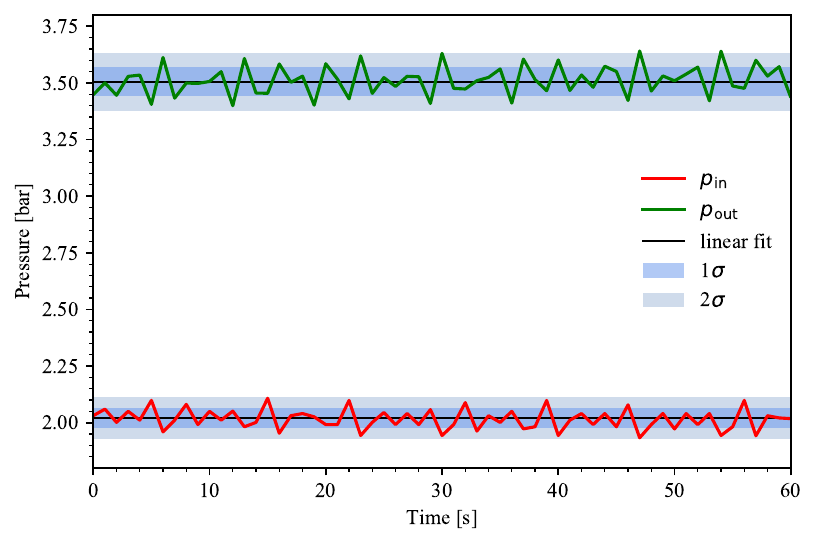}

\caption{\label{fig:6}Four cylinder performance distribution of inlet pressure $p_{\text{in}}$ (red) and outlet pressure $p_{\text{out}}$ (green) with xenon gas at a not shifted movement over a time span of 60\,sec. A linear regression curve and the 1\,$\sigma$ and 2\,$\sigma$ confidence band are shown.}
\end{figure}

\begin{figure}[htbp]
\centering 
\includegraphics[width=0.95\textwidth]{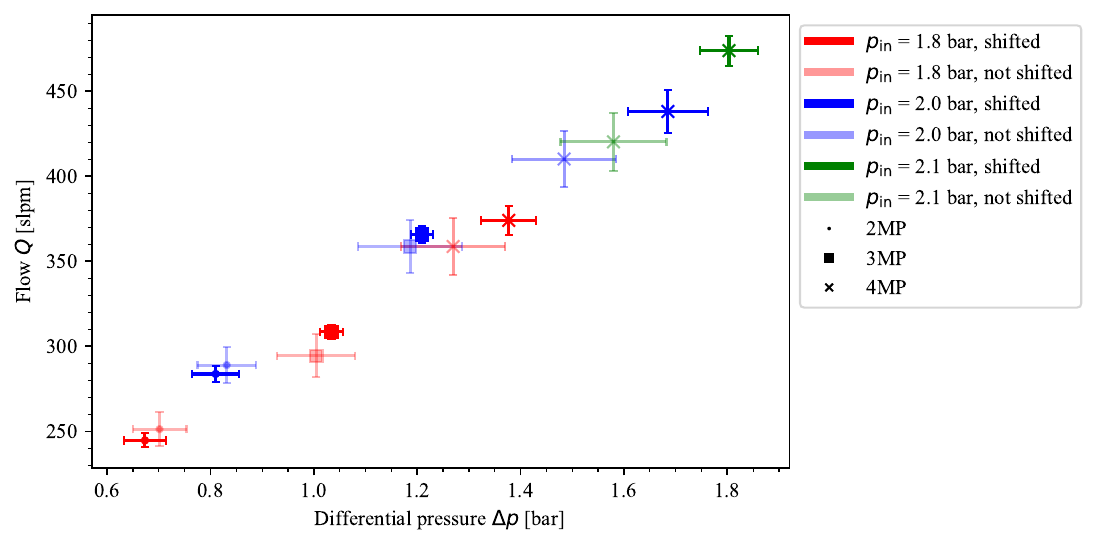}

\caption{\label{fig:7}Four cylinder performance with xenon gas in the closed loop mode. The flow $Q$ as a function of the differential pressure $\Delta p$ was measured for several configurations with two (2MP, point), three (3MP, square) and four pumps (4MP, cross) at different inlet pressures $p_{\text{in}}$. Each measurement was done once with a phase-shifted movement and once with a not shifted movement.}
\end{figure}

Figure\,\ref{fig:6} illustrates the pressure distribution of the inlet pressure $p_{\text{in}}$ and the outlet pressure $p_{\text{out}}$ for the not shifted operation with all four cylinder pumps over a time span of 60\,sec. The 1\,$\sigma$ and 2\,$\sigma$ bands indicate the pressure fluctuations. Such Fluctuations on the size of up to $\pm$90\,mbar for $p_{\text{in}}$ and up to $\pm$130\,mbar for $p_{\text{out}}$  need to be reduced by the synchronized operation with a phase-shifted movement ensuring a stable operation of connected systems like the radon removal system. In order to quantify the performance at several configurations and the shift-related improvement, the performance was measured with different configurations for the multi-piston mode, namely two (2MP), three (3MP) and four pumps (4MP). The results are visualized  in figure\,\ref{fig:7}. 

To quantify the reduction of fluctuations in flow and differential pressure, each measurement was done once with the optimal shift for each configuration and once without shift (not shifted).

Except for the 2MP configuration, the optimal shift is not only reducing the fluctuations, but also giving an additional gas flow and differential pressure boost. For the 3MP a flow increase up to 5\% and a differential pressure increase up to 3\% could be achieved by the phase shifted movement. The fluctuation could be reduced by up to 32\% in differential pressure and 21\% in flow. The 4MP performance was boosted in flow by 13\% and in differential pressure by 14\% in maximum. Its fluctuations were reduced by up to 52\% in flow and 55\% in differential pressure.

At an inlet pressure of 2\,bar corresponding to the operation pressure of the radon removal system, the 2MP configuration at optimal shift (blue, point) reaches a gas flow of ($284 \pm 5$)\,slpm at a differential pressure of ($0.81 \pm 0.05$)\,bar.
The 3MP configuration reaches at similar conditions (blue, square) a flow of ($366 \pm 5$)\,slpm with a differential pressure of ($1.21 \pm 0.02$)\,bar.
With the 4MP configuration, a flow of ($438 \pm 13$)\,slpm and a differential pressure of ($1.68 \pm 0.08$)\,bar were achieved (blue, cross).

The best 4MP performance at optimal shift was achieved at an inlet pressure of 2.1\,bar, the maximum inlet pressure operated in this work: A flow of ($474 \pm 9$)\,slpm at a differential pressure of ($1.80 \pm 0.06$)\,bar was measured. Higher inlet pressures are expected to lead to an even higher performance.

\begin{figure}[htbp]
\centering 
\includegraphics[width=0.8\textwidth]{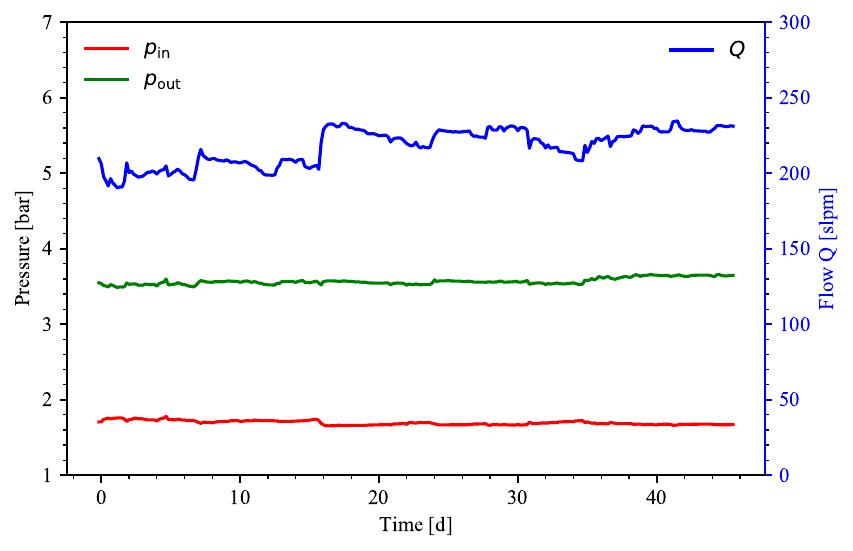}

\caption{\label{fig:8} Four cylinder pump long-term performance acting as compressor for the XENONnT radon removal system. The pump inlet pressure $p_{\text{in}}$ (red) during operation of the distillation column at an operational pressure of 2\,bar was decreased to 1.7\,bar due to the flow resistance of the lines between both systems. The created pump outlet pressure $p_{\text{out}}$ of 3.5\,bar (green) is large enough for the liquefaction of the related xenon gas flow $Q$ (blue) in a heat exchanger of the distillation column. Several distillation modes were tested over the long-term period of 46\,days indicated by the change in flow over time.}
\end{figure}

Figure\,\ref{fig:8} shows the four cylinder long-term performance acting as a compressor during the commissioning of the XENONnT radon removal system. The four cylinders were operated with the optimal phase shift. The pump inlet pressure $p_{\text{in}}$ of 1.7\,bar (red) is given by the operation pressure of the distillation column reduced by the flow resistance induced pressure drop of the connection between both systems. The pump outlet pressure $p_{\text{out}}$ (green) measured at the collective line after the outlet buffer volume of each cylinder pump of 3.5\,bar is required to liquefy the related xenon gas flow (blue) in a heat exchanger of the distillation column. During the 46\,day long operation, several distillation modes were tested explaining the flow changes over time between 200\,slpm and 235\,slpm xenon. Thus, the four cylinder magnetically-piston pump fulfills all requirements for the operation of the XENONnT radon removal system.

\newpage
\section{Conclusion}\label{IV}

An ultra-clean radon-free four cylinder magnetically-coupled piston pump was developed, built and successfully commissioned to be operated as a xenon gas compressor in an energy-efficient heat pump at the novel radon removal system of the dark matter experiment XENONnT.

The pump concept is based on a prototype pump \cite{e} developed for the gas purification systems of the experiments XENON1T and nEXO. A few design parameters,
like diameter, length, flow resistance and gasket material,
were optimized for the application at XENONnT to meet the requirements of pressure difference between in- and outlet, flow and cleanliness.  
The usage of only clean and radio-pure materials lead to a measured average radon emanation of ($75 \pm 13$)\,$\upmu$Bq for a single cylinder pump.
A new key feature is the use of four cylinder pumps connected in parallel to increase the flow. They feature a phase-shifted synchronization of their movements to not only boost the performance but also to reduce flow and output pressure fluctuations.
The custom-made programming of the synchronization gives the possibility to operate the system with different configurations and to monitor the status of each pump during the operation.

The performance of each individual pump was measured with xenon gas for several inlet pressures. Additionally, the multi-cylinder performance was measured for different phase shifts and different number of cylinder pumps. A maximum performance of $Q=(474 \pm 9)$\,slpm flow at a differential pressure of $\Delta p = (1.80 \pm 0.06)$\,bar was measured for a xenon inlet pressure of $p_{{\text{in}}} = 2.1$\,bar. A long-term operation of 46\, days during the commissioning of the radon removal system of XENONnT was successfully performed at the required design parameters.

Overall, its high, stable and long-term performance combined with low oscillations regarding flow and pressure, the special cleanliness and the radio-purity make this pump system interesting for current and future rare event experiments using not only xenon but also other noble gases.

The concept presented here allows for scalability in terms of dimension of each individual cylinder defining the achievable pressure difference $\Delta p$ and flow $Q$ (see equations (\ref{eq:fource}) to (\ref{eq:flow})) and in terms of the number of pumps connected in parallel. In combination with the screening and decision on nearly all components used, this leads to a variety of possible applications.
In principle, a pump based on this concept could be used in  projects beyond rare event searches which require a displacement pump for ultra-clean gases or even liquids, for example in the medical or industrial sector.

\newpage\clearpage

\acknowledgments
We gratefully acknowledge support from the group of M. Lindner at MPIK Heidelberg by determining the radon emanation of all four cylinder pumps and several piston gasket materials. Furthermore we express special thanks to the Electrical and the Mechanical Workshop of the Institut f\"ur Kernphysik at M\"unster University.
The pump development at Muenster University have been supported by BMBF (05A20PM1).

\typeout{}
\bibliographystyle{achemso}

\end{document}